\documentstyle[12pt,aaspp4]{article}
\def\IZw18{I~Zw~18}

\def\deg{$^{\circ}$~}
\def\msun{${\rm ~M_\odot}$~}
\def\zsun{${\rm ~Z_{\odot}}$~}

\def\mmsun{~{\rm M_\odot}~}
\def\msunyr{$~{\rm M_\odot}$~yr$^{-1}$~}

\def\angs{~\AA~}
\def\Ha{H$\alpha$~}
\def\Hb{H$\beta$~}

\def\n{NGC~}
\def\asec{\ifmmode {'' }\else $''~$\fi}  
\def\amin{\ifmmode {' }\else $'~$\fi}    
\def\sles{\lower2pt\hbox{$\buildrel {\scriptstyle <}
   \over {\scriptstyle\sim}$}} 
\def\sgreat{\lower2pt\hbox{$\buildrel {\scriptstyle >}
  \over {\scriptstyle\sim}$}} 
\def\s2{[SII]~$\lambda\lambda6717,31$}
\def\kms{~km~s$^{-1}$~}
\def\ergsec{~ergs~s$^{-1}$~}

\def\flux{~ergs~s$^{-1}$~cm$^{-2}$}
\def\cm3{~cm$^{-3}$}
\def\fig{{Figure}}
\def\x{{X-ray}~}

\def\et{{\rm et\thinspace al.}\ }   
\def\apj{ApJ}
\def\apjs{ApJS}
\def\pasp{PASP}
\def\aj{AJ}
\def\mn{MNRAS}

\def\aa{A\&A}
\def\aasup{A\&AS}

\def\aar{A\&AR}
\def\pasj{PASJ}
\begin{document}

\title{Kinematic Evidence for Superbubbles in \IZw18:  Constraints on the
Star Formation History and Chemical Evolution}

\author{Crystal L. Martin\altaffilmark{1,}\altaffilmark{2}}

\altaffiltext{1}{Visiting Astronomer, Kitt Peak National Observatory, NOAO,
operated by AURA, Inc, under contract with the NSF.}

\altaffiltext{2}{Observations reported here have been obtained in part with
the Multiple Mirror Telescope, a joint facility of the University of Arizona
and the Smithsonian Institution.}

\affil{Steward Observatory, University of Arizona, Tucson, AZ 85721}

\begin{abstract}
We have combined measurements of the kinematics, morphology, and 
oxygen abundance of the ionized gas in \IZw18, one of the most
metal-poor galaxies known, to examine the  star formation history 
and  chemical mixing processes.
Deep \Ha imagery shows diffuse emission and a partial shell
extending well beyond the main two knots of
 continuum emission.
We have explored the kinematics of this ionized gas using
 longslit, echelle spectroscopy of the \Ha line.
We find the unambiguous
signature of a supergiant shell southwest of the galaxy and weak evidence 
for a second bubble northeast of the galaxy.
The axial symmetry of these shells and the asymmetry in their \Ha line profiles
suggest that they comprise the lobes of a single bipolar bubble expanding
at $\sim 30 - 60$~km~s$^{-1}$.
Higher velocity gas is found near the small shell immediately west
of the northwest HII region.
Although an unresolved \x  source is discovered near the northwest  HII region
 in  archival ROSAT PSPC data, we argue that  
hot interstellar gas associated with the superbubbles does not produce all
the \x emission.
Oxygen emission lines are detected up to
$\sim 1$~kpc from the NW HII region along the bubble's polar axis,
so this diffuse, ionized gas has been polluted with gas processed
by stars.
Measurements of the  O/H  abundance ratio  in the inner nebula show 
surprisingly little variation considering the apparent youth of the
galaxy.

We describe the dynamical evolution of the superbubble
using a  simple wind-blown bubble model.
To test the hypothesis that the dynamical age of the bubble measures
the duration of the starburst in \IZw18, 
we compute the photometric properties of a starburst with the same
age as the superbubble.
We find that star formation commencing
$15 - 27$~Myr ago and continuing at a rate of 
0.017 -- 0.021 \msun (of 1 -- 100\msun
stars) per year can both power the gas dynamics and produce a fair
match to the integrated
optical properties of \IZw18.
The total mechanical energy returned to the interstellar medium by stellar
winds and supernovae,
7~--~30~$\times 10^{53}$~ergs, is  insufficient to eject the
entire interstellar medium.
However,  the corresponding mechanical energy injection rate 
is high enough to drive  the superbubble  shell   out of the HI gas cloud, and
``blowout'' will allow the hot ISM to escape in a galactic wind.
This supports the idea that metal-enriched winds play a prominent role
in the chemical evolution of dwarf galaxies.

\end{abstract}

\keywords{galaxies: individual {\IZw18} -- ISM: kinematics and dynamics --
galaxies: evolution -- galaxies: abundances}

\section{Introduction}
\label{sec:intro}

The blue compact dwarf galaxy, \IZw18, was first described as
``a double system of compact galaxies'' having an emission line
spectrum with a fairly featureless continuum (Zwicky 1966).
Its extremely blue colors and exceptionally low oxygen abundance 
indicate the present star formation rate 
exceeds the past average rate  (Searle \& Sargent 1972).
Kunth \& Sargent (1986) have argued
that the metal abundance in \IZw18 --
O/H $\approx$ 0.02 (O/H)$_{\odot}$ in the HII regions
(Skillman \& Kennicutt 1993 and references therein) --
 is close to the minimum observable
in any self-enriched, HII region.
The prospect that \IZw18 may be forming stars for the first
time out of primordial clouds of gas has drawn much attention
(e.g. Sargent \& Searle 1970; 
Lequeux \& Viallefond 1980; Kunth, Lequeux, \& Sargent 1994), and 
 the galaxy has played a prominent role in determining
the primordial He abundance 
(Lequeux \et 1979; Davidson \& Kinman 1985; Pagel \et 1992).


The question of whether \IZw18  formed stars prior to the current burst
is not settled, but several recent results support the young galaxy hypothesis.
The colors extracted from UBV imaging studies of blue compact dwarfs typically uncover
an older stellar population underlying the starburst.  However,
comparable efforts for \IZw18 show no evidence for an old population
  (Sudarsky \& Salzer 1995).
Broadband HST images do resolve several spatially distinct 
populations within the main body of \IZw18, but again  
show no evidence for an older population
(Hunter \& Thronson 1995).
The three continuum patches strung out to the
NW of the main body have progressively redder colors and are probably associated
with the galaxy (Davidson, Kinman, \& Friedman 1989; Dufour \& Hester 1990).
Dufour \& Hester  have speculated that  the elements synthesized  
in these earlier, isolated star formation events
escaped without significant mixing into the HI clouds.
The C, N, and O abundances of the main HII regions are well described by
a single burst  that started about 10~Myr ago and
 restrict models with previous star formation 
to a short burst at least  $10^9$ years ago (Kunth, Matteucci, \& Marconi 1995).
Measurements of the metallicity of the HI gas 
(Kunth, Matteucci, \& Marconi 1995; Pettini \& Lipman 1995) 
should distinguish between these chemical evolution models, whose
success relies on some primary production of N and a differential
galactic wind (Marconi, Matteucci, \& Tosi 1994).



Although the structure of the \Ha and 21-cm emission has been
described previously (Dufour \& Hester 1990; Viallefonde \et 1987), 
several important questions about the gas dynamics remain.
For example, Skillman \& Kennicutt (1993) have questioned
the assumption that the HI velocity field reflects a rotating
disk.  The implied total mass is 13 times larger than the HI mass
(Viallefond \et 1987) and causes \IZw18 to deviate significantly
from the mass-metallicity relation observed for dwarf irregular
galaxies (Skillman \et 1988).
Examining the velocity field on finer scales
will determine the contribution of non-virial motions  to the large-scale
gas motions.
Another outstanding question is the role of a galactic wind.
Meurer (1991) has suggested  that the \Ha emission extending roughly
perpendicular to the main HI cloud is a minor axis outflow.
It is important to substantiate this claim with supporting kinematic
evidence.  



We selected \IZw18 for a case study within a broader program investigating
the interplay between star formation and the ISM (Martin 1996).
The feedback could have a  particularly strong influence on the
evolution of  \IZw18 since the low
metallicity reduces radiative losses 
and the  escape velocity is relatively low.
This paper presents 
new  imagery and  longslit echelle spectra of the \Ha
emission  and previously unreported \x emission
(\S~\ref{sec:observe}).
The objectives are to describe the kinematics of the ionized gas 
(\S~\ref{sec:observe}),
derive a consistent dynamical interpretation (\S~\ref{sec:models}),
determine the implications for  the star formation history of \IZw18
 (\S~\ref{sec:models}), 
and  estimate what effect  mass loss is likely to have on the galaxy's evolution
(\S~\ref{sec:discuss}).  
Since the results suggest superbubbles mix the interstellar gas on kiloparsec scales,
we examine new optical spectra in  \S~\ref{sec:chem} and  place limits
on the O/H abundance variations in the extended ionized gas.
The results are summarized in \S~\ref{sec:sum}.

\section{Observations and Analysis}
\label{sec:observe}
\subsection{Optical Imaging}
\label{sec:imaging}
\begin{figure}
\epsscale{0.65}
\plotone{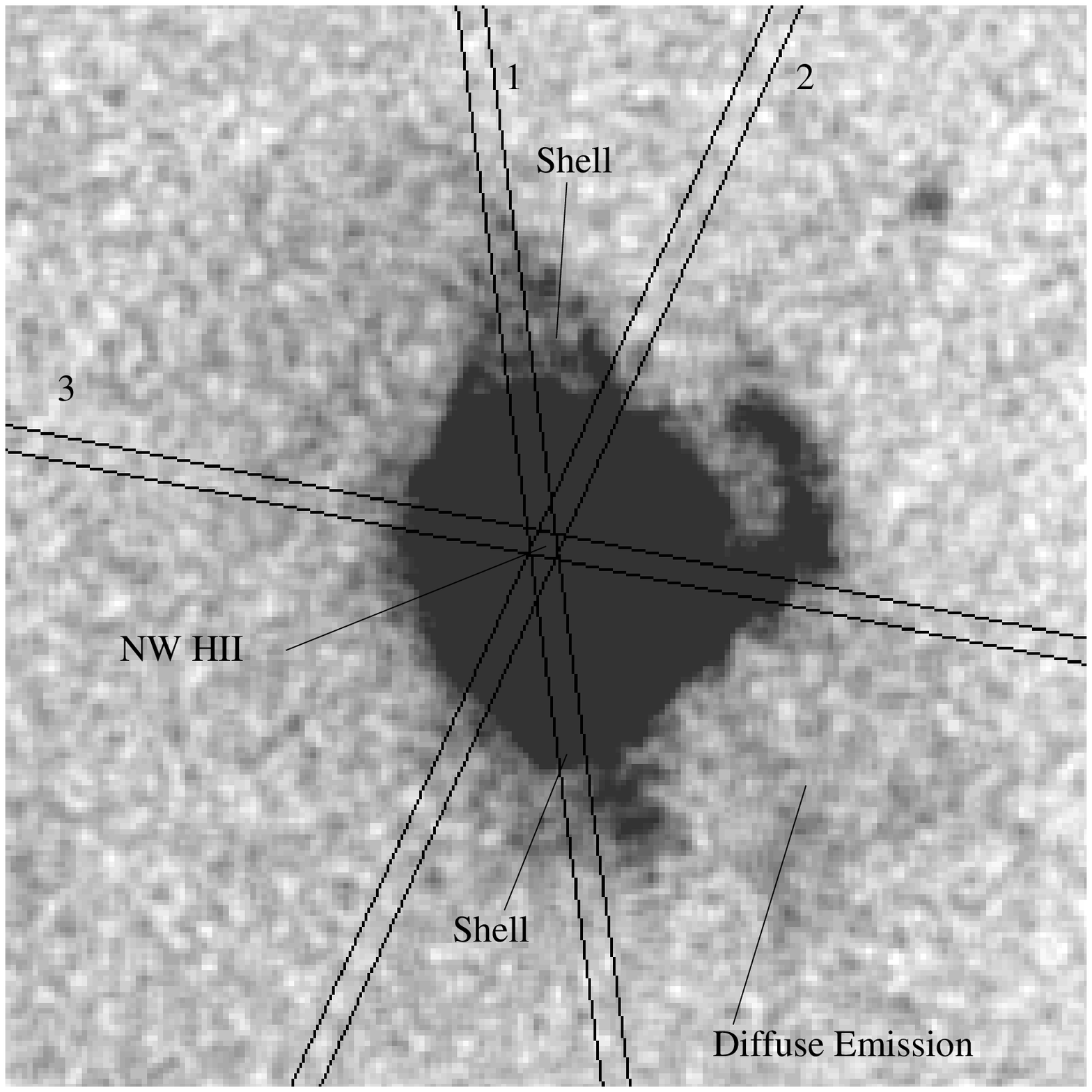}
\caption{
(a) \Ha image of \IZw18  with
 echelle slit positions  overlaid.
(b) Same as `a' but the intensity is displayed on a logarithmic scale to
show the structure of the inner nebula.}
\plotone{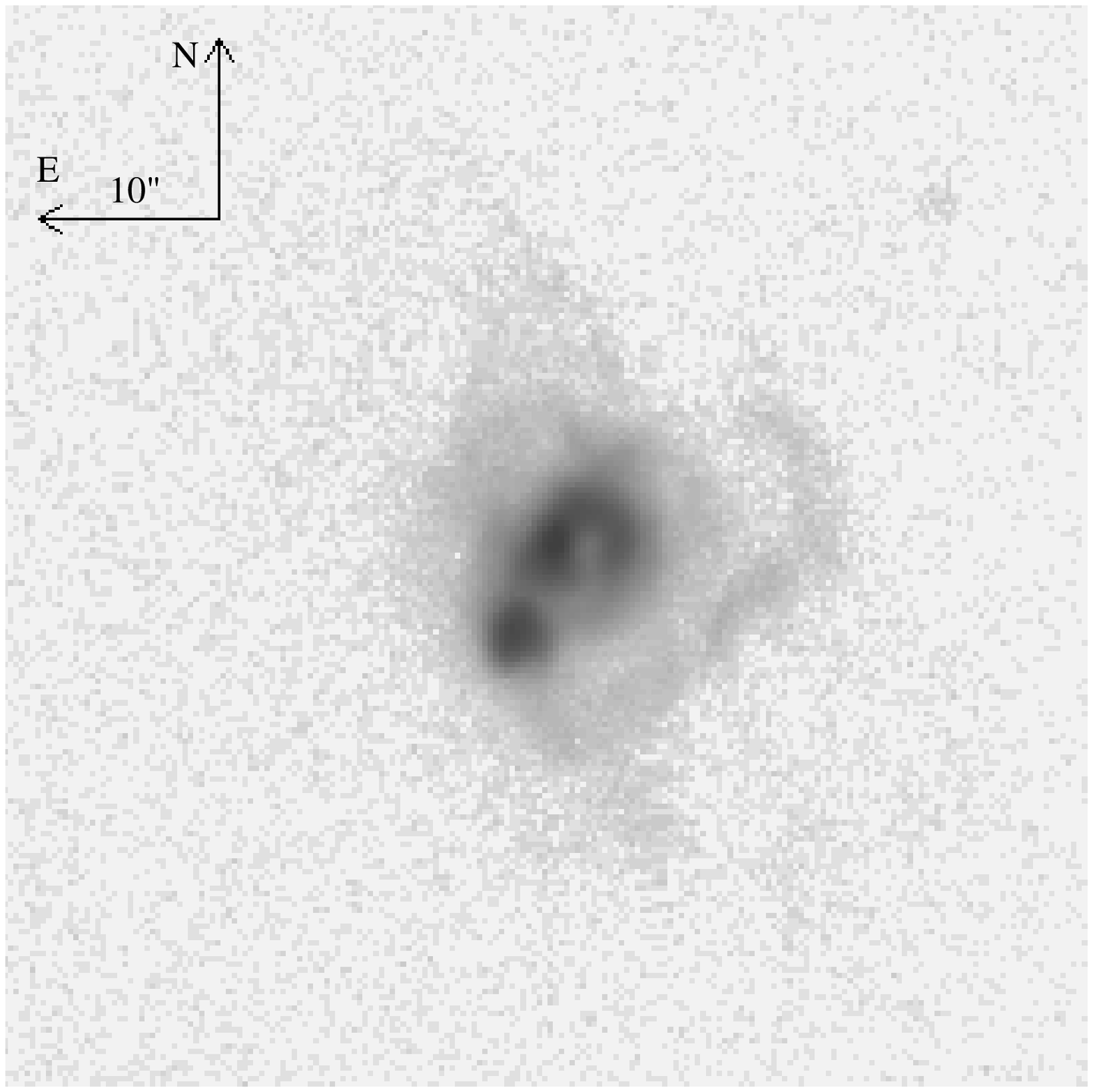}
\label{fig:outer}
\end{figure}

Narrowband \Ha and red continuum images of \IZw18 were obtained 1993 May 17 with the
Steward Observatory 2.3~m telescope equipped with a Loral $800 \times 1200$ thinned
CCD.   The raw CCD frames were processed with standard techniques as 
described in Martin \& Kennicutt (1995).
Figures~\ref{fig:outer}ab show the continuum subtracted \Ha image.
The  NW HII region is centered 
about 1\asec E of the  brightest continuum emission.
A small, partial shell of diameter 3\farcs6 (175 pc)
\footnote
{A distance of 10~Mpc to \IZw18 is adopted throughout this paper
(Viallefond, Lequeux, \& Comte 1987).}
 protrudes from the 
NW side of the NW HII region and wraps around the continuum emission
(Davidson \et 1989; Hunter \& Thronson 1995).
The SE HII region is coincident with the second brightest maximum 
 in the continuum emission.
The cores of both HII regions are offset NE of the two maxima in the HI map
(Dufour \& Hester 1990).
The deeper reproduction in \fig~1a  reveals additional
 features absent in the off-band image.  For example,
the ridge to the SW is thought to represent a 
radiation-bounded ionization front being driven into the main HI cloud
(Dufour \& Hester 1990).
A prominent shell stretches  15\asec (740 pc) N-NE from the NW HII region, and
bright \Ha emission   extends symmetrically  S-SW of the NW HII region.
Faint, diffuse  emission 
(emission measure $EM \approx 50$ pc~cm$^{-6}$ for $T_e = 1.9 \times 
10^4$~K)
is detected along a  position angle of  42\deg in a band 21\asec (1 kpc) wide
extending 
25\asec (1.2 kpc)  and 31\asec (1.5 kpc) to the NE and  SW, respectively.
The 21-cm emission is less extended in this direction;
the principal 30\asec by 60\asec HI cloud 
is elongated along the optical major axis,
PA~$\approx$~ 328\deg (Viallefond \et 1987).


The \Ha image was flux calibrated using the absolute flux reported by 
Dufour \& Hester (1990) through a
1\amin $\times$ 1\amin square aperture centered on the NW HII region.
Absolute fluxes were corrected for atmospheric extinction and foreground Galactic
reddening ($A_V$ = 0.21 mag);  the contribution of [NII] lines is negligible.
The integrated \Ha flux within 30\asec of the NW HII region is
$5 \pm 1 \times 10^{-13}$\flux, where the 20\% uncertainty is dominated by the flux
calibration.
Assuming the nebula is radiation-bounded, 
 Case~B conditions,  and  an electron temperature $T_e = 1.9 \times 10^4$~K
(e.g. Skillman \& Kennicutt 1993), the  flux of hydrogen ionizing
photons is $1.4 \pm 0.2 \times 10^{52} ~d_{10}^2$~s$^{-1}$, 
where $d_{10}$ is the distance to \IZw18  in units of 10~Mpc.
This ionizing luminosity is
similar to that of 30~Doradus in the Large Magellanic Cloud (Kennicutt 1984).
A spherical, homogeneous nebula of this size would have
 an rms electron density of $n_{\rm rms} = 0.39 \pm 0.04 ~d_{10}^{-1/2}$~cm$^{-3}$.
The ratio of the \s2 line fluxes is near the 
low density limit, so the actual electron density in the emission line filaments
is $n_e < 250$\cm3. The volume filling factor of the ionized
filaments must be at least $\epsilon > 1.0 \times 10^{-3}$, where
$\epsilon \equiv n_{rms}^2 / n_e^2$.
The estimated mass of ionized gas is then
$M_{HII} = 1.0 \pm 0.1 \times 10^7 \mmsun \sqrt{\frac{\epsilon}{0.01}}
d_{10}^{5/2}$.
To estimate the radial  density profile, the \Ha surface brightness  was 
azimuthally averaged in  annuli around the NW HII region.
Assuming the nebula is spherical,
inversion of the \Ha surface brightness integral implies
$n_{\rm rms} = n_{e} \sqrt{\epsilon} \propto r^{-1}$ within 7\asec (340~pc)
 of the NW HII region, and $n_{\rm rms} = 
n_{e} \sqrt{\epsilon} \propto r^{-2}$ between 7\asec  and 15\asec (730~pc).

\subsection{High-Resolution Spectroscopy}
\label{sec:echelle}
Longslit spectra of the \Ha line were obtained 1994 April 29 and 30 
using the echelle spectrograph on the KPNO 4~m telescope.
Three position angles were chosen based on the morphology of the \Ha emission
 (see \fig~\ref{fig:outer}).
An astrometric offset was used to center the slit  on the NW HII region,
 and three 20 minute integrations were obtained at each position angle.
The 1\farcs5 slit produced a spectral resolution of $\sim 11$\kms FWHM.  
The instrumental setup, calibration techniques, and data reduction 
are further described by Martin and Kennicutt (1995).  
The seeing varied from 1-2\asec FWHM, so
spectra were extracted every 1\farcs8 along the slit to maximize the spatial
information.
This binning provided sufficient S/N 
to measure the line center to 
an accuracy of 0.1 \AA~ and fit two-component Gaussian line profiles 
down to the faintest intensities visible in \fig~\ref{fig:outer}a.

\begin{figure}
\epsscale{0.75}

\plotone{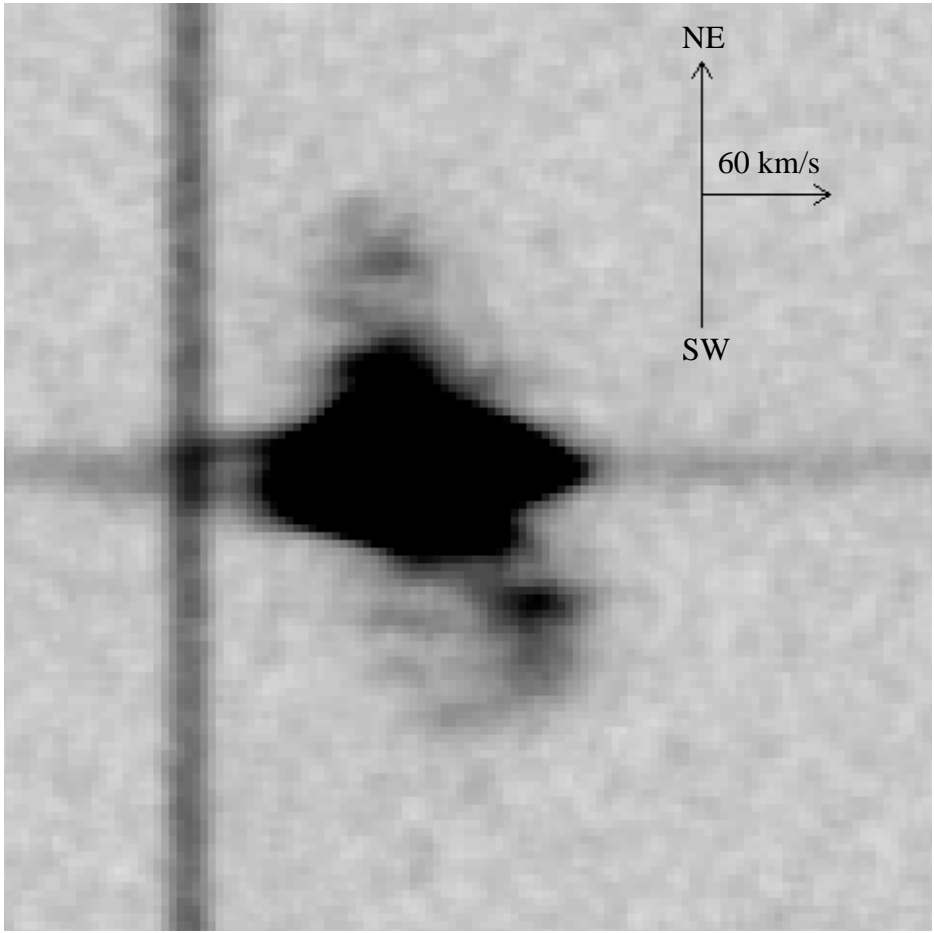}
\caption
{Echelle spectrum of \Ha emission line  along slit 1.  
The horizontal line is the continuum of the NW HII region.
A prominent Doppler ellipse is visible to the SW, and  a
 second Doppler ellipse is discernible to  the NE.
Blue wings from  high velocity gas near the HII region  extend 
past the night-sky emission line  (vertical line).
The vertical arrow is approximately 20\asec long.
}
\label{fig:slit1}
\end{figure}

The echellograms reveal the kinematic signatures of two expanding supergiant shells.
Figure~\ref{fig:slit1}  shows the echellogram along slit 1, PA = 7.7\deg. 
Southwest of the 
NW HII region (continuum source), the line profile splits into two components
forming a Doppler ellipse.
Similar kinematic signatures have been observed in other starbursting,
gas-rich dwarfs and almost certainly indicate the presence of an expanding shell
of gas (Marlowe \et 1995; Martin 1996).
This Doppler ellipse  extends 19\asec (930~pc) to the SW, and the 
maximum separation of the line-of-sight velocities  reaches  59\kms.
Fits to the line profile northeast of the HII region are improved with a
faint, secondary component redward of the main component, thereby
providing some evidence for
a second Doppler ellipse coincident with the NE shell in the \Ha image.

\begin{figure}
\epsscale{1.0}
\plotone{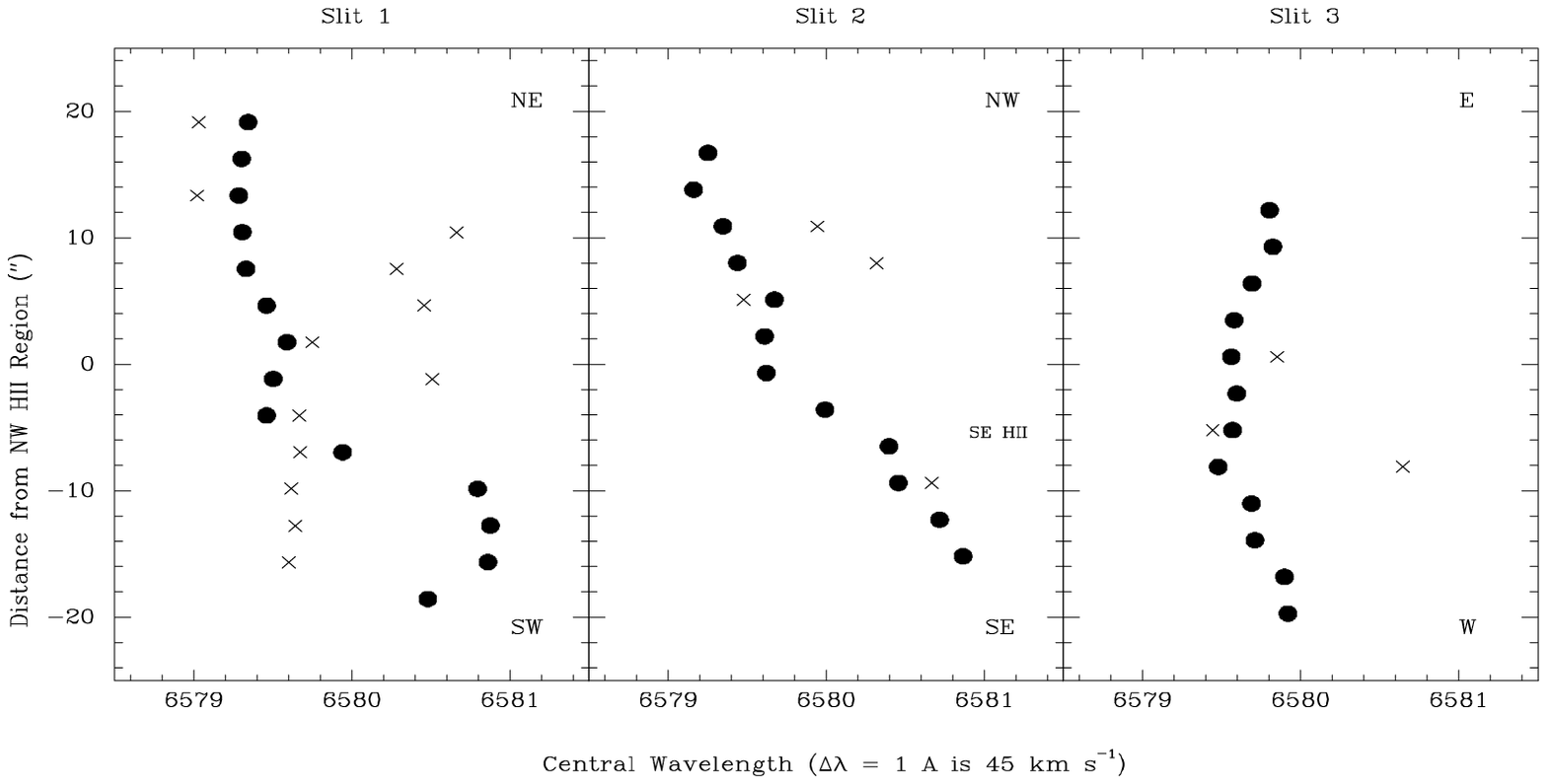}
\caption[ha position-velocity diagram]
{\Ha position-velocity diagrams.
The solid circles mark the central velocity of the brighter
\Ha component, and the weaker component is denoted by an `X'.  
The double component fit was adopted at positions where it significantly  
decreased the rms residuals relative to a single component fit.  
Slit positions are described in the text and shown in \fig~\ref{fig:outer}a.
}
\label{fig:ha_pv}
\end{figure}

The  asymmetry in the line profiles of  these shells
suggest the superbubbles form  the two lobes of a bipolar bubble
with polar axis inclined relative to our line-of-sight.
In the  NE Doppler ellipse the intensity of the blueshifted component is 
several times higher than that of the redshifted component, but
their relative intensities are reversed to the SW.
This effect is illustrated  in \fig~\ref{fig:ha_pv}, where the symbol
``X'' denotes the position of the  weaker component in position-velocity space.
The lack of line-splitting along the other slit positions  lends  credence to the
bipolar bubble interpretation.

We also detect gas with velocities up to $\pm 230$\kms from
the systemic velocity and tentatively associate it with the small shell.
These high-velocity wings are stronger on the blue side of the line profile and
 coincident with the small shell along all three slit positions.
However, since this high-velocity  component contributes only  5\%
of the line flux, the only other region with high enough signal-to-noise
to reveal these wings is the SE HII region; and examination of the data along
slit~2 shows a similar wing contributes 3\% of the \Ha flux there.
We hesitate to identify the high-velocity gas with an expanding shell because
no Doppler ellipse is seen where slit~3 crosses the center of the small shell.
A champagne flow (e.g. 
Yorke, Tenorio-Tagle, \& Bodenheimer 1984) from the young HII regions 
and stellar winds might provide a better description of the gas kinematics.

The detection of large, expanding shells of gas reopens the question of
whether galactic rotation produces all the velocity shear across the gas
distribution.
In \fig~\ref{fig:ha_pv} the solid circles trace the velocity of the dominant
component of the \Ha emission along each of the three slit positions.
Along slit 1, the velocity is fairly constant to the NE but increases 
by  $\sim 73$\kms to the SW.
At least part of this  velocity shear is likely an
artifact of the expansion of the superbubble.
Along slit 2, PA = 156.1\deg, 
the velocity gradient is steeper on the SE side of the nebula.
Since this slit position is nearly parallel to the HI major axis, it might be 
expected to reveal the galactic rotation curve.
However, the change in slope near the NW HII region and the similarity 
to the velocity gradient along slit~1
leave some ambiguity  between the effects of galactic rotation
and the expansion of the superbubble.
The velocity is nearly constant along 
slit~3, PA = 78.8\deg, which is 24\deg from the HI minor axis.




Along all three slit positions, the large-scale variations in the
central velocity of the \Ha line appear to be shared by
the HI velocity field (Viallefond \et 1987).
Given the better spatial coverage of the HI
map and the strong suggestion of galactic rotation in it,  we
provisionally attribute the  velocity gradient across echelle slit~2 
to galactic rotation.
Higher resolution HI observations are needed to determine whether
the kinematic signatures of the bubbles  are present in the HI 
and what affect they might have on the global HI velocity field.
A significant perturbation on the local velocity field is already evident
in the NE finger of HI, which is coincident  with our large \Ha shell.  
The iso-velocity contours there depart from the overall pattern,
and the HI radial velocity decreases by about
20\kms across the  \Ha shell.


\subsection{X-Ray Detection}
\IZw18 was observed  with the Position Sensitive Proportional Counter (PSPC) on the
R\"{o}ntgensatellit (ROSAT) for 16,964 sec during 1992 April 30 - 1992 May 11.
We requested  the data, originally obtained by Dr. C. Motch, from the ROSAT public
archive.
An  unresolved \x source is located
 2\farcs6 W and 6\farcs6 S of the peak red continuum emission.
Pointing errors of this magnitude are typical for ROSAT observations, and we identify
the \x source with \IZw18.  It is not clear whether a second, fainter source
located 82\asec to the NNW is associated with \IZw18.
After background subtraction, the  remaining $\sim 130$ net counts are insufficient to 
constrain a spectral fit.  The \x colors, however,
suggest a fraction of the \x flux is produced by hot, coronal gas in the ISM.  
The emission is  slightly harder than the soft emission from the dwarf
\n5253 (Martin \& Kennicutt 1995), but  
considerably softer than the hard  sources, thought to be massive \x binaries, 
near the center of \n2403  (Martin 1996).
To estimate the 
\x luminosity, $L_x$, in the ROSAT band (0.1 -- 2.2 keV), a series of 
power law, bremsstrahlung, and Raymond-Smith spectral
models were normalized to the total counts.  For a Galactic absorbing column
$N_{HI} ~\sles~ 2 \times 10^{20}$~cm$^{-2}$ (Heiles 1975; Stark \et 1992),
all these models  produce $L_x \approx 1 \times
10^{39}$\ergsec to within a factor of two.
For example, a Raymond-Smith model for a $5 \times
10^6$~K plasma of 10\% solar abundance with a foreground absorbing column of
$\log N_{HI} = 20.2$ implies $L_x = 1.1 \times 10^{39}$~ergs~s$^{-1}$.

Until ROSAT HRI observations are obtained, the extent of the source will 
remain unknown.
At 1~keV the FWHM of the ROSAT/PSPC point spread function is 24\asec, about
half the angle subtended by the \Ha emission along slit~1. However, the
radius encircling 95\% of the photons is 
substantially larger -- about 45\asec (Hasinger \et 1992) -- so a diffuse
halo should not be ruled out.
If the \x source is found to reside deep within the galaxy, the intrinsic
absorption column could be as high as $N_{HI} = 3.5 \times 10^{21}$~cm$^{-2}$
(Kunth \et 1994);
and the  estimated \x luminosity would be increased
 by a factor 
$ < 7$, where the upper limit is derived for a column with solar abundances.


%
%

\section{Modeling the Gas Dynamics}
\label{sec:models}
In this section,  a simple model  for a wind-blown bubble
is used with the kinematic and morphological properties
described in \S~\ref{sec:observe}  to constrain
the dynamical age of the superbubble and the mechanical energy supplied to it.
The luminous properties of a starburst which could
drive the expansion are computed and compared to 
the integrated properties of the galaxy.


\subsection{The Superbubble Model}
\label{sec:scope}

\begin{figure}
\epsscale{0.75}
\plotone{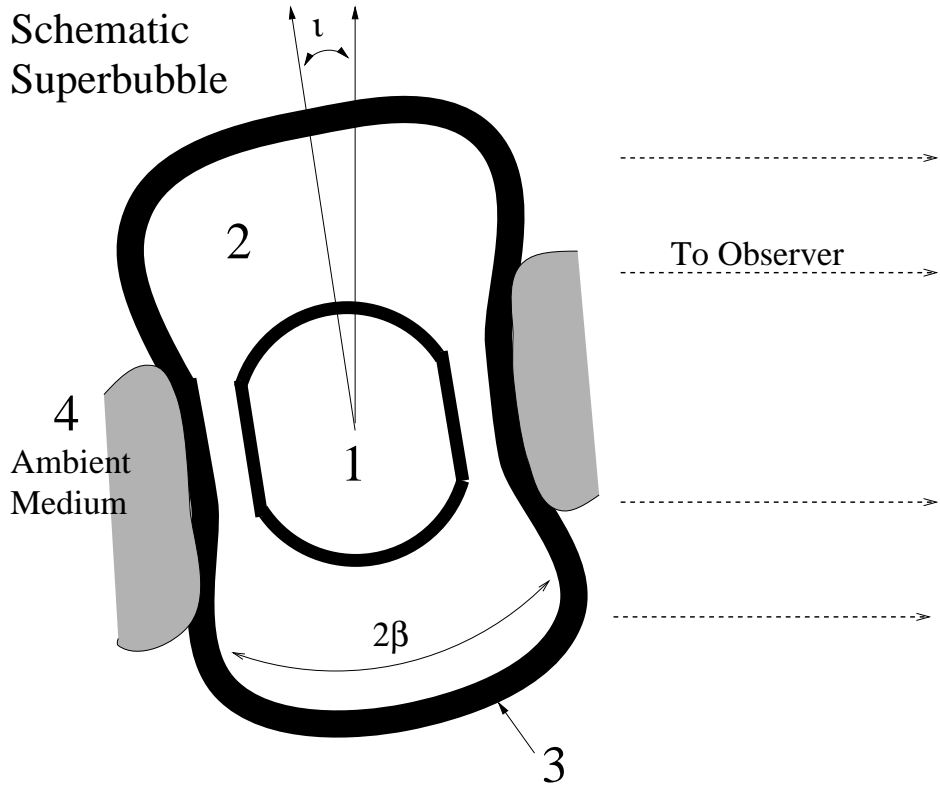}
\caption[schematic]
{Cross section of favored bipolar bubble geometry.
Heavy, solid lines represent the shocked shell, zone 3, of swept-up ambient
medium, zone 4,  and the stand-off shock
separating the hot bubble (zone 2)  and the starburst (zone 1).
The polar axis is inclined $\iota \sim 10$\deg to our line-of-sight, and 
the angle
subtended by a lobe from the central starburst is $\beta \sim 30$\deg.
}
\label{fig:cone}
\end{figure}

The dynamical arguments presented in this paper are based on the standard
model for a stellar wind bubble with a radiative shell 
(e.g. Castor, J., McCray, R., \& Weaver, R. 1975; Weaver, McCray, \& Castor 1977).
The conceptual framework
is illustrated schematically in \fig~\ref{fig:cone}.
Stellar winds and supernova explosions from
hundreds of massive stars fuel a supergiant bubble over timescales 
$\sgreat~ 10^7$~yr (zone 1).  The  kinetic energy in the ensemble's supersonic wind
is thermalized by a stand-off shock, and the high pressure (zone 2) downstream drives
a strong shock into the ambient ISM (zone 4).  The swept-up gas condenses into a 
shell  
(zone 3) as a result of radiative cooling.

The solution for the shell's evolution as  formulated by  Ostriker \& McKee (1988) 
provides a convenient power-law parameterization of a density gradient
in the ambient medium,
\begin{equation} 
   \bar{\rho} = \bar\rho(1) \left[ \frac{R_s}{R_s(1)} \right] ^ {-\kappa_{\rho}},
\end{equation}
and a time-dependent energy injection rate,
\begin{equation}  \label{eqn:L}
  L_{in} = L_{in}(1) \left[ \frac{t}{t(1)} \right] ^ {\eta_{in} - 1}.
\end{equation}
In this notation,  $\bar{\rho}(1)$ is the average density of the ambient
ISM within the fiducial radius $R_s(1)$, and
supernovae and stellar winds supply a mechanical power
$L_{in}(1)$ at time t(1).
The parameters t(1) and $R_s(1)$ are not independent and must satisfy
the relation
${R_s} / {R_s(1)} = \left( {t} / {t(1)} \right) ^ \eta$, where
$\eta \equiv (2 + \eta_{in}) / (5 - k_{\rho})$.


The radius of the spherical swept-up shell is 
\begin{equation}    \label{eqn:r}
R_s(t, L_{in}(1) / \bar{\rho}(1)) = R_s(1)  
 \left( \frac{\xi \Gamma L_{in}(1)}{\eta_{in} \bar{\rho}(1) R_s(1)^5} 
\right) ^ {\eta / 3} t^{\eta},
\end{equation}
where $\xi$ and $\Gamma$ are numerical constants on the order of unity
as defined by
Ostriker \& McKee (1988) and evaluated for a  ratio of specific heats $\gamma = 5/3$.
The shell velocity is 
\begin{equation}  \label{eqn:v}
v(t,L_{in}(1) / \bar{\rho}(1)) = \frac{dR_s}{dt} =
\eta R_s(1)  
 \left( \frac{\xi \Gamma L_{in}(1)}{\eta_{in} \bar{\rho}(1) R_s(1)^5} 
\right) ^ {\eta / 3} t^{\eta - 1}.
\end{equation}
Solving  equations \ref{eqn:r} and \ref{eqn:v}  
gives the dynamical age of the bubble,
\begin{equation}
 t = \eta \frac{R_s}{v}, 
\end{equation}
and the ratio of the mechanical power to ambient density, 
\begin{equation} \label{eqn:ln}
 \frac{L_{in}(1)}{\bar{\rho}(1)} = \frac{\eta_{in}}{\xi \Gamma \eta^3}
R_s(1) ^{(5\eta - 3)/ \eta} v^3 R_s^{3(1 - \eta)/ \eta}. 
\end{equation}
The total mechanical energy injected into a bubble of age $t$ is
\begin{equation}
 E_{in}(t) = \frac{L_{in}(1) t(1)}{\eta_{in}} \frac{t^{\eta_{in}}}{t(1)^{\eta_{in}}}.
\end{equation}
%


\subsection{Application to \IZw18}
\label{sec:geo}

The Doppler ellipse discovered SW of the HII regions is not the signature
of a spherical shell expanding radially.
The measured radius and velocity would require a burst of star formation
in the center of the bubble about 6~Myr ago which supplied
$E_{in} = 1.8 \times 10^{53} \bar{n}(1)$~ergs of kinetic energy.
The continuous emission from the fading population would easily
be detected in our red continuum image.

We favor interpreting  the data in the context of
an expanding bipolar bubble.
The  bright \Ha emission coincident with the SW shell
shares a common symmetry axis with
the NE loop, which may form the second lobe.
This polar axis intersects the brightest region of the galaxy --
the obvious place for a power source.
The  reversal  of the shape of the \Ha line profile between the 
two bubbles also supports this interpretation since it is easily
explained by tilting the bubble with respect to our line-of-sight.
Overall, the kinematic signature is   reminiscient of
that along the minor axis of M82, which exhibits the quintessential superwind
(Heckman, Armus, \& Miley 1990; Martin 1996).
However, while the outflow in M82 is clearly confined at its waist
by the galactic disk, the collimating mechanism in \IZw18 is not apparent.
The polar axis presumably depicts the projection of the steepest pressure gradient,
which we might expect to be perpendicular to the HI major axis.  The proposed polar
axis is 43\deg from the PA of the HI major axis in the map of Viallefond \et  (1987).
We note that the major axis of the very diffuse, extended \Ha emission 
is  perpendicular to the HI major axis.

To estimate the radius and expansion speed of the SW lobe from the data,
we introduce two simplistic models for the shape and velocity field of the bipolar
shell.
A cone of half-opening angle $\beta$ and
inclination $\dot{\iota}$, as illustrated in \fig~\ref{fig:cone}, represents
the thick shell behind the outer shock.
We assume the density decreases as $r^{-2}$, where $r$ is the distance from the
starburst, which is consistent with the decline in surface brightness in our
\Ha image.
The inclination is not large because the shifts of the red and blue components
of the SW Doppler ellipse with respect to the central velocity are similar.
We estimate $\beta \sim 30$\deg from the width of the NE loop.
An inclination of $\iota \sim 10$\deg then produces an intensity contrast of
2 -- 3, similar to the data. The radial extent of the Doppler ellipse is 
essentially the length of the polar axis.
In our ``isobaric bubble'' model, the shell velocity is normal to the conical surface,
geometrical corrections are small, and we estimate an expansion speed $v \sim 35$\kms.
(The variations in the magnitude of the velocity around the shell is thought to 
be considerably smaller.)
In our ``wind'' model,  the flow is radial, mimicking the boundary layer
between a freely expanding supersonic wind and the halo, and the deprojected 
shell speed is $v = 61$\kms.
Hence, the shell velocity is uncertain by a factor of 2. The uncertainty in
the radius is dominated by the difference, about 20\%, between the two lobes.

Despite the complex geometry, we  apply the  dynamical model  of \S~\ref{sec:models}
along the polar axis. We expect to estimate the dynamical age to within a factor
of two and $E_{in} / \bar{n}(1)$ to better than an order of magnitude based, primarily,
on the  uncertainties in radius, velocity, and density gradient ($k_{\rho}$) rather
than the choice of dynamical model.  The magnitude of the errors are believed to be similar.
For example,
 the height of the polar lobe is larger than the radius of a spherical bubble
supplied with the same energy, but  a radial  density gradient could also
 easily increase $R$ by a factor of two
relative to the constant density case.
Our best estimate of the dynamical age from the isobaric bubble and wind geometries
are 27~Myr and 15~Myr, respectively.  The bubble requires $E_{in} = 
7.5 \times 10^{54} ~\bar{n}(1)$~ergs
of kinetic energy while the wind geometry increases the estimate to $E_{in} = 
2.2 \times 10^{55} 
~\bar{n}(1)$~ergs.


The mean density of the ambient ISM swept up by the shell is expected to be low.
If all the HI gas were compressed
 into a uniform sphere the size of the bipolar bubble ($R_s \sim 900$~kpc), the
average density would be $\bar n(1) = 0.63$\cm3, where the number density
$\bar{\rho}(1) = \bar{n}(1) m_H \mu$ and
$\mu = 1.4$~amu is the average mass per H atom for 
 a composition H:He = 10:1 by number.
Adding the density of ionized H in the  nebula (\S~\ref{sec:imaging}) 
 yields an upper limit
$\bar n(1) < 1$\cm3 for the $k_{\rho} = 0$ models.
Spreading the HI  out uniformly over the HI clouds suggests
a lower limit of $ 0.019$\cm3 $ ~<~  \bar n(1)$.
To explore the effect of a 
density gradient  on  galactic scales, we calculated models with
 $k_{\rho} = 1$~and~2 and $R_s(1) = 100$~pc.
By an analogous argument, the acceptable ranges for the ambient density
interior to $R_s(1)$ are
 $ 0.57 ~\sles~  \bar n(1) ~\sles~ 3.1$\cm3 and
 $ 5.3 ~\sles~  \bar n(1) ~\sles~ 15$\cm3,
for $k_{\rho} = 1$~and~2  respectively.
These central densities are similar to estimates of the HI  density  in the center 
of the main cloud ($\sim 10$\cm3) and the rms electron density in the NW HII region.





%
\subsection{The Star Formation History}
\label{sec:sb}


Either a recent starburst or an IMF heavily biased towards massive stars
can produce colors as blue as observed for \IZw18.  
Searle \& Sargent (1972) ruled out the latter interpretation for \IZw18 
with a measurement of the gas phase O/H abundance ratio.
Recently  the stellar content of 
the starburst has  been studied using very high spatial resolution.
Stars down to O9.5 on the main
sequence are resolved in broad-band, HST images (Hunter \& Thronson 1995), 
and  Hunter \& Thronson identify three spatially distinct populations.
First, the stars inside the small shell have ages from 1-- 5~Myr.
The second  population in the southern component has no red supergiants (RSGs)
and is probably younger.
The third population is spread throughout the galaxy,
comprises roughly half of the resolved stars,
and is likely to be older than the shell population since it contains more red stars.
In addition to this  ``general galaxy'' population, an older, unresolved
population with the colors of B or early A stars follows the main body
of the galaxy and has an age $\sgreat~ 10^7$~Myr.

Our detection of high velocity  gas near the small shell confirms
the observations of Davidson \et (1989) and supports the conclusion of 
Hunter \& Thronson (1995) that the shell population powers  the  small shell.
As discussed in \S~\ref{sec:echelle}, 
our data do not unambiguously determine the expansion velocity of the shell.
So, it remains unclear whether a few a massive stars that formed early created the
shell, or whether a younger bubble is driven by many stars.

The supergiant bubble is clearly older than the shell
cluster.  Could its dynamical age indicate the duration of the 
starburst in \IZw18~?  
To  check the  consistency of this hypothesis, we use 
the evolutionary synthesis models of Leitherer \& Heckman (1995) to
describe the photometric properties of a stellar population which 
could power the bubble.
We adopt  their  lowest metallicity models, $0.1Z_{\odot}$, and 
consider only an IMF of slope $\alpha = 2.35$, $M_{up} = 100$\msun, and 
$M_{low} = 1$\msun.

An instantaneous burst model is not a good description of the star formation
history of \IZw18.  A 15~Myr old population that produced  hydrogen
ionzing photons at the observed rate would generate 300 times more blue
luminosity than observed.
A continuous star formation rate  over the $10-30$~Myr lifetime of the
bubble seems reasonable (cf. Meurer \et 1995)
and is consistent with the detection of
stars only a few million years old (Hunter \& Thronson 1995).
Up to an age $\sim 40$~Myr,
the mechanical power, $L_{in}(t)$, generated by this burst will grow with time as shown
in \fig~56 (lower right) of Leitherer \& Heckman (1995), and we incorporate
this evolution in our dynamical model by setting $\eta_{in} = 3$ and
$t(1) = 40$~Myr.  The steep increase in the wind power keeps the shell moving
at a constant velocity, which would also be obtained from  
a $k_{\rho} = 2$ density gradient.

The continuous star formation model produces a stellar population whose
photometric properties resemble \IZw18.
As shown in Table~2, this result is not very sensitive to our choice
of geometry as represented by the ``isobaric bubble'' and ``wind'' models.
At  dynamical ages of 15~Myr and 27~Myr,  star formation rates of 
0.021 \msunyr (of 1 to 100 \msun stars) and 0.017 \msunyr, respectively,
 are required to generate the blue luminosity of \IZw18.
These model populations produce only half of the  ionizing luminosity
 measured and are a bit redder than \IZw18.  This discrepancy is not
bothersome as roughly 20\% of the \Ha flux comes from the compact
source on the eastern edge of the small shell.  We could speculate that
this  star formation is an isolated, triggered event or that the global
star formation rate is increasing. This population might be  too young and hot
to contribute substantially to the blue luminosity.

The continuous burst models easily produce enough supernovae and stellar 
winds to drive the superbubble.
After 15~Myr and 27~Myr, the kinetic energy imparted to stellar ejecta is
$6.6 \times 10^{53}$~ergs and $2.7 \times 10^{54}$~ergs, respectively.
Since our echelle data constrain the ratio $E_{in} / \bar{n}(1)$ rather
than $E_{in}$ alone, we can choose the ambient density, $\bar{n}(1)$, to
make our dynamical model consistent with the  starburst model of the same age.
The inferred densities, 0.03\cm3 and 0.36\cm3, are within the range estimated
from the HI data in \S~\ref{sec:geo}.

\subsection{Results}

The aim of our modeling exercise was to describe the range of star formation
histories that could easily explain the gas kinematics in \IZw18.
Uncertainties about the geometry of the superbubble and the  distribution
of the ambient gas limit the accuracy of our analysis.  We stepped through
the analysis for two models representing the likely range of 
shell velocities.  We found that the star formation necessary to power the
superbubble also produces most of the starburst's luminosity.
This  epoch of star formation probably started 15 -- 27 Myr ago.
We favor a higher age within this range because our ``isobaric bubble''
model is probably the closer analogy to the actual shell kinematics and
also predicts a volume-averaged ambient density closer to expectations.

The evolutionary tracks for low metallicity, massive stars are controversial 
(e.g. Renzini \et 1992) and could introduce errors
in the models for the young starbursts.
Near an age of 4~Myr,
the blue luminosity of the 0.1\zsun instantaneous burst
brightens 65\% more than  the solar
metallicity models (\fig~9 of Leitherer \& Heckman 1995).
Eliminating this jump would increase the ratio of ionizing luminosity
to blue luminosity ($Q/L_B$).
Since the error diminishes with age and an increase in  $Q/L_B$ would 
actually improve the agreement with the observations, 
a small error of this nature will not change our  conclusions.

\section{Discussion}
\label{sec:discuss}
In this section, we use our results from \S~\ref{sec:models}
 to explore the influence of the superbubble on the evolution of \IZw18.
A particularly interesting issue is the amount of global mass loss
driven by the burst of star formation.
It has been argued on a theoretical basis that a global wind created by
the first burst of star formation may eject the interstellar gas from a
dwarf galaxy (e.g. Larson 1974; Saito 1979; Dekel \& Silk 1986; Vader 1986).
The transformation of gas-rich dwarfs to dwarf ellipticals by this mechanism
is appealing but is not completely consistent with the systematic structural
properties of dwarfs (Ferguson \& Binggeli 1995, and references therein).
It has also been emphasized that the escaping wind should be enriched
with metals recently dispersed in Type~II supernovae (Vader 1987; DeYoung \&
Gallagher 1990).  These
differential galactic winds have  been independently invoked to 
explain the locus  of blue compact and dwarf irregular galaxies
 in  the
N/O - O/H and He/H - O/H planes (Marconi, Matteucci, \& Tosi 1994).
In \S~\ref{sec:massloss}, we discuss the conditions necessary for the bipolar
bubble to develop into a galactic wind, the escape of the hot gas 
from the galaxy's gravitational grip, and the mass of cooler gas permanently ejected.
In \S~\ref{sec:chem}, we discuss the mass of metals produced by our
starburst models and the role of the superbubble in mixing
these elements into the interstellar gas.
In \S~\ref{sec:pilot},  measurements of O/H abundance variations
are presented and  discussed.

\subsection{A Starburst-Driven Galactic Wind}
\label{sec:massloss}

\subsubsection{Development of an Outflow}
The power requirement for the superbubble to break free of the ISM
is characterized by the minimum
 mechanical power, $L_P$, necessary for the bubble to grow
as large as the gaseous extent of the galaxy.
Following the derivation of Koo \& McKee (1992), 
in which the effective scale height is defined by
\begin{equation}
 H_{eff} \equiv \frac{1}{\rho_0} \int_0^{\infty} \rho(z) dz ,
\end{equation}
the critical rate of kinetic energy injection is
\begin{equation}
 L_P = 17.9 \rho_0 H_{eff}^2 c_{s0}^3 {\rm ~ergs~s}^{-1},
\end{equation}
where $c_{s0}$ and $\rho_0$ are the isothermal sound speed 
and midplane gas density in the ambient medium.
Since $c_{s0}$ is intended to serve as a measure of the pressure in
the ISM, we interpret $c_{s0}$ as the effective one-dimensional velocity
dispersion of the HI gas. The velocity dispersions measured in galactic disks
typically fall in the narrow range from 3 -- 10\kms (Kennicutt 1989), 
consistent with estimates of 8\kms for the Galactic disk (McKee 1990).
The HI linewidth 
provides  an upper limit of $\sigma = 19$\kms (Kunth \et 1994).
From  HI observations (Viallefond \et 1987; Lequeux \& Viallefond 1980),
we estimate a scale height $360~{\rm pc} ~\sles~ H_{eff} < 720$~pc.  
In units of  $c_{s0} = 13$\kms and $H_{eff} = 360$~pc, the
 ``breakout''  threshold is
\begin{equation}
 \frac{L_P}{n_0} = (1.13 \times 10^{38} {\rm ~ergs~s}^{-1}~{\rm cm}^{3})~
 H_{360}^2 c_{13}^3,
\end{equation}
where $n_0 = \rho_0 / (\mu m_H)$ is the  midplane number density.
The  HI density in the core of the main cloud provides an upper 
limit of $n_0 \sim  9$\cm3.
Normalized to the blue luminosity of \IZw18,
our isobaric bubble and wind models in
 \S~\ref{sec:models} were supplied with kinetic energy at  time-averaged rates 
of
 $\bar{L_{in}(t)} = 3.2\times 10^{39}$\ergsec and
$ 1.4\times 10^{39}$\ergsec 
after 27~Myr and 15~Myr, respectively.
We conclude that the bipolar bubble will likely
break through the HI layer supersonically.

To open a channel for the hot gas inside the bubble to flow out of the galaxy, 
the shell must accelerate and  break up from Rayleigh-Taylor instabilities.
Mac Low \&  McCray (1988) and Mac Low, McCray, \& Norman 
found that  the wind luminosity required
for this  ``blowout'' to occur is larger than $L_P$.
Their numerical simulations  suggest that $L_{in} / L_P ~\sgreat~ 5$ is a sufficient 
condition for  blowout.
Large $L_{in} / L_P \sim 1000$ produced blowout 
when the shell reached a height  $\sim 3 H_{eff}$; but,
for smaller $L_{in} / L_P$, 
they suspect the bubble may grow larger in
the z-direction before blowing out.
For $n_0 = 1$\cm3,
 the ratio of the superbubble's mechanical power to the characteristic
luminosity is $L_{in} / L_P = 12 - 28$, and we expect the superbubble to
 blowout.

\subsubsection{Gravitational Potential}
The mass of gas that will escape from the galaxy following blowout depends
on the depth of the gravitational potential well.
We assume the mass distribution is similar to an isothermal sphere
and use the circular velocity,
$ v_c^2 = r {d\phi}/ {dr}$, to estimate the depth of the potential well.
For a halo that extends to $r_{\rm max}$, 
the escape velocity is given by
\begin{equation}
v_{\rm esc} = \sqrt{2} v_c \sqrt{1 - \ln(r/r_{\max})}.
\end{equation}
Using Bernoulli's theorem, we find
the  corresponding escape temperature,
\begin{equation}
T_{\rm esc} = \frac{\gamma - 1}{2\gamma } 
\frac{\mu m_H}{k}v_{\rm esc}^2, 
\end{equation}
which describes the specific thermal energy required to establish
a smooth, supersonic outflow.

If the velocity gradient along the HI major axis of \IZw18
is produced by a rotating disk, then the deprojected circular
velocity is  $v_c \sim 40$\kms. 
The dynamical mass is then $\sim 9 \times 10^8$\msun (Viallefond \et 1987),
and halos truncated at 1~kpc and 10~kpc,  masses of 
  $3.7 \times 10^8 \mmsun$  and $3.7 \times 10^9 \mmsun$ respectively,
provide representative models.
For reference, 
the escape velocity and escape temperature are tabulated in Table~3
as a function of radius.


\subsubsection{Mass Loss}
\label{sec:ml2}
The predicted temperature  for the bubble interior,  $ \sim 5 \times10^6$~K
(from Eqn.~4 of Martin \& Kennicutt 1995), 
is  considerably higher than
the escape temperatures shown in column~4 of Table~3, so
the hot bubble interior (zone 2) should easily escape following blowout.
The ensuing question is
how much mass resides in  this hot phase of interstellar gas.
In the canonical superbubble model, the hot interior 
is composed of shocked stellar ejecta and material
conductively evaporated off the swept-up shell.
From Eqn.~9 of Shull (1993), we estimate that 
the evaporated mass is $\sim (4 - 11) \times 10^5 \mmsun$, 
which is more than the
$\sim (2 - 5) \times 10^4$\msun of enriched gas returned to the ISM
by the starburst models.
Clouds that get run over by the shell and evaporate in the bubble's hot interior  may further mass load the bubble
(Martin \& Kennicutt 1995).

Unfortunately, the \x observation provides only a rough upper limit on the
mass of hot gas.  
If   all the \x flux comes from hot gas of density $n_x$,
the observed  luminosity implies a mass of coronal gas
$M_x \sim (1.1 \times 10^6 \mmsun)(0.05~{\rm cm}^{-3}/n_x) $ 
for a plasma with cosmic abundance.
(Although the value of $n_x$ is unknown, arguments about the gas pressure
in the ISM suggest it is unlikely to exceed 0.05\cm3.)
At a lower metallicity, 
the emissivity is expected to be lower, so the inferred mass would be
even higher.
This analysis merely demonstrates that thermal emission from hot gas
could make a substantial contribution to the \x luminosity.


Although some of the gas in the shell may escape from the galaxy with
the hot wind, the  starburst  will probably not eject the entire ISM
of \IZw18.
For example, the escape velocity
from the lower mass halo in Table~3 is 
$\sim 60$\kms  near the shell radius.
Since  the deprojected shell velocity is probably $30 - 60$\kms, only
fragments of
the ruptured shell could coast out of the galaxy.
However, the starburst
would have to transfer $\sim 7 \times 10^{54}$~ergs of kinetic energy
to the interstellar  gas to accelerate all the HI to
100\kms,  a typical escape velocity in  Table~3. 
In the spherical superbubble model,
only 20\% of the injected mechanical energy is converted into
the shell's kinetic energy, so the starburst  would have to generate
at least  $4 \times 10^{55}$ ergs of  mechanical energy
to completely eject the ISM.
In \S~\ref{sec:sb}
we estimated  values of $E_{in}$ 
 over an order of magnitude smaller than this, so the
 complete ejection of the ISM is not energetically possible.

\subsection{Chemical Enrichment of the ISM}
\label{sec:chem}

Our dynamical interpretation of the gas kinematics has two interesting
 implications for the chemical enrichment of \IZw18.
First, elements synthesized and ejected by the starburst's massive stars may
have been transported  $\sim 900$~pc across the bubble interior
 in only 15 -- 27~Myr.
Second, most of the oxygen produced may not reside in the warm, ionized gas --
a puzzle independently pointed out by others (Kunth \& Sargent 1986).
Using a fit to the  oxygen yields calculated for 12 -- 100 \msun stars
(Prantzos 1994), we derive an oxygen yield $y = 0.01$ for an IMF with lower
and upper mass limits of 0.1\msun and 100\msun and slope $\alpha = 2.35$, 
where $y$ is the ratio
of the net mass of oxygen produced to the total mass permanently locked up in
remnants (e.g. Tinsley 1980).
The starburst models in \S~\ref{sec:sb} formed $(3 - 5) \times 10^5$~\msun of 
1 -- 100 \msun stars, or about $(8 - 12) \times 10^5$~\msun of 0.1 -- 100 \msun stars,
so $9 \pm 3 \times 10^3$ \msun of oxygen are returned to the interstellar gas.
If all this oxygen was mixed into the warm, ionized gas, the metallicity
of the HII regions would be 5 -- 6 times higher than measured.
The  oxygen would have to be mixed throughout the entire mass of HI
to produce a  homogeneous   oxygen abundance as low as 2\% $(O/H)_{\odot}$.
This unlikely scenario contradicts our hypothesis that the current burst 
is the first major star formation episode and that the size of the bubble is an 
indication of the mixing scale.
We suggest that the extra oxygen may still be in the hot phase of the interstellar
gas -- zone~2 in \fig~\ref{fig:cone}.
Assuming a remnant mass of 1.5\msun for the stars in the 12 -- 100 \msun range,
we estimate that 10\% of the mass turned into stars has been returned to the ISM.
The oxygen mass fraction of this gas is then  $X_O \approx 9 ~X_O^{\odot}$, 
where the solar value
is  $X_O^{\odot} = 9 \times 10^{-3}$ (Grevesse \& Noels 1993).
Since gas evaporated from the cooler phases of the ISM will dilute
the hot stellar ejecta (\S~\ref{sec:ml2}), we expect oxygen to 
comprise $ ~\sles~ 0.65 - 1.6 \%$ 
of the mass of  hot interstellar gas (i.e. $X_O ~\sles~ 0.7 - 1.7 ~X_O^{\odot}$).





\subsection{The O/H Abundance Ratio in the Ionized Gas}
\label{sec:pilot}

If  the superbubble plays a prominent role in the dispersal of metals 
over large scales in \IZw18, significant abundance  inhomogeneities might 
be expected (cf. Kunth \et 1994). We have examined
deep longslit, optical spectra of \IZw18, obtained for a related project
(Martin 1996),
to search for an abundance edge to the metal-enriched region and to constrain
the chemical homogeneity of the ionized gas. 


\subsubsection{Optical Spectrophotometry}

Longslit optical spectra of \IZw18 were obtained in 1994 on the MMT
using the Blue Channel Spectrograph equipped with  a
 Loral 3k $\times$ 1k CCD detector.
A 500 gpm grating blazed at 5410\angs in 1st order
was used with  a UV-36 blocking filter and a 1\asec slit.
This configuration provides  spectral coverage from approximately 3700\angs
to 6800\angs  at a moderate spectral
resolution,  about $\sim 5$~\angs FWHM.
Spectra were obtained at two slit positions, one
 centered on the NW HII region and rotated to  PA 7.6\deg (slit a) and
another positioned across
the SE HII region at a PA of 131.0\deg   (slit b); the total integration times
were 12,000 s and  9,900 s, respectively.
The spectrum of the night sky background was  recorded on both
ends of the slit.
The seeing limited the resolution along the slit to $\sim 2\asec$.
Care was taken to monitor the effects of varying parallactic angle over the
time sequence of frames (Filippenko 1982).


The data reduction followed standard techniques and employed the IRAF
\footnote{
IRAF is distributed by the National Optical Astronomical Observatories, which are
operated by the Association of Universities for Research in Astronomy, Inc. (AURA),
under contract with the National Science Foundation.}
 software package.
The raw CCD frames were bias subtracted, corrected for pixel-to-pixel sensitivity
variations and slit illumination, transformed using exposures of an
HeNeAr arc lamp, and  extinction corrected.  A new field flattener was being
tested during the November run, and these data required an additional distortion
correction.
Cosmic rays were removed when the individual frames were combined.
The spectra were flux calibrated
using observations of standard stars (Massey et al. 1988).

\subsubsection{Primordial Ionized Gas ?}
\begin{figure}
\epsscale{0.75}
\plotone{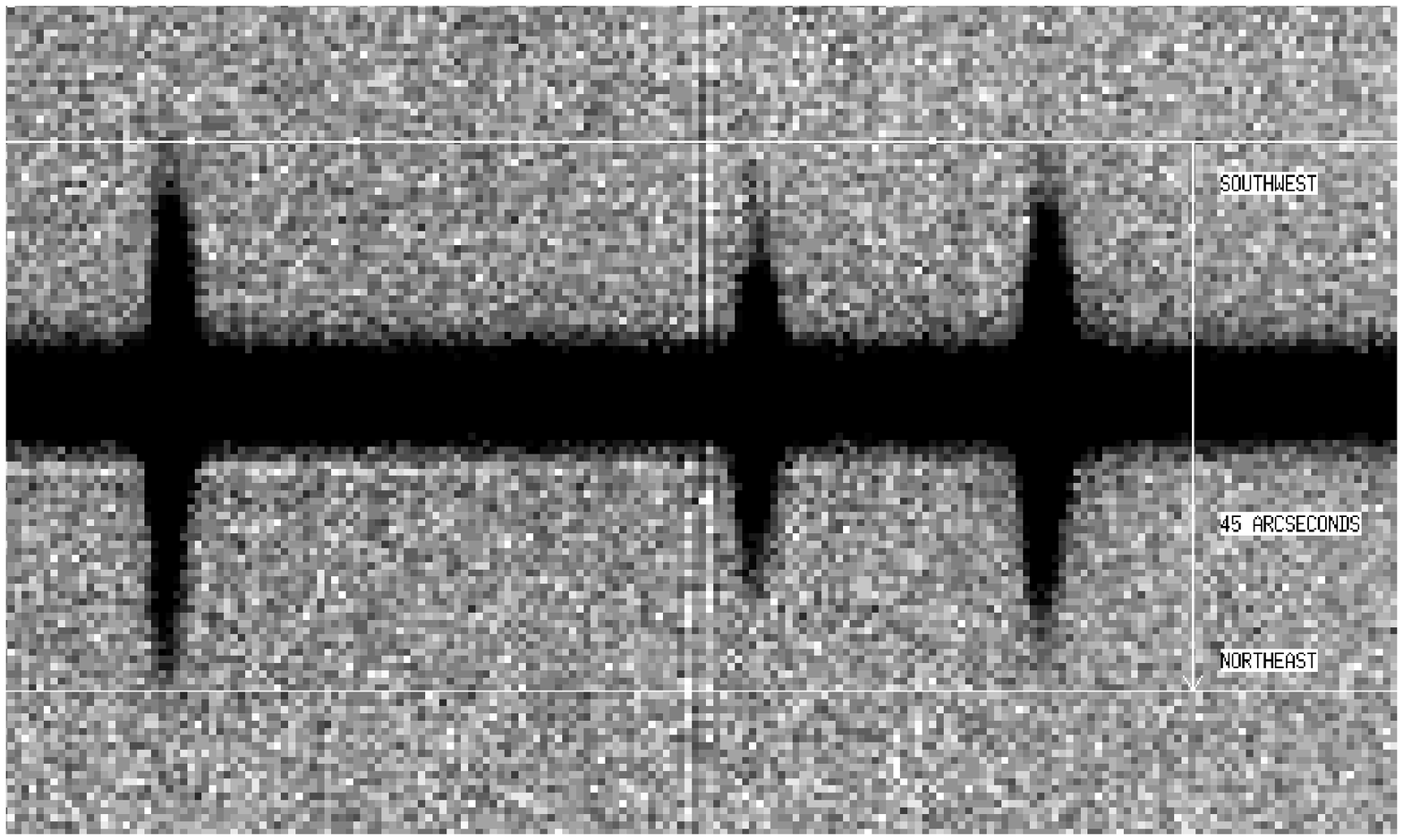}
\caption[mmt]
{Longslit spectrum along the polar axis of the superbubble.
The continuum, running horizontally, is centered on the NW HII region.
The  \Hb emission line is on the left and 
 [OIII] $\lambda \lambda 5007, 4959$ are on the right.
}
\label{fig:mmt}
\end{figure}

\fig~\ref{fig:mmt}
 demonstrates that the ionized gas across the region where the superbubbles
are detected is not primordial.  The [OIII] $\lambda5007$
emission extends along slit ``a''  as far (23\asec or 1100~pc) to the SW
as the \Hb emission line which has comparable intensity.
To the NE the intensity of [OIII]  falls faster than the \Hb intensity but
is clearly detected to 21\asec (1030~pc).
A lower O abundance to the NE or a lower ionization state could explain
the declining [OIII] $\lambda5007$ / \Hb ratio.
Along slit~``b'', the emission lines [OIII]~$\lambda5007$ and \Hb are spatially
coincident and  detected over  35\asec.
However, we find no evidence 
for an abrupt edge to the O-enriched gas within the ionized ISM in \IZw18,
as might be expected if the galaxy contains an ultra-low metallicity HI
halo (Kunth \et 1994).

\subsubsection{Abundance Variations}

\begin{figure}
\epsscale{0.85}
\plotone{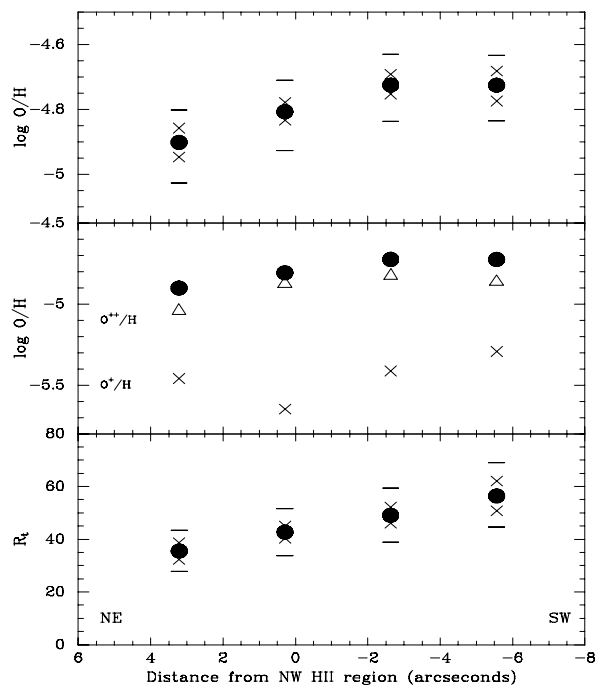}
\caption[oh]
{Abundance variations along slit~a.
(a) Temperature ratio, $R_t \equiv$~[OIII] $\lambda \lambda 5007, 4959
/ \lambda 4363$.
(b) Ionic abundance ratios and their sum.
(c) O/H abundance ratio with error estimates as described in the text.
}
\label{fig:oh}
\end{figure}

We use the data along slit~``a''
to place an upper limit on the variation of the O/H
abundance  ratio within the inner region of the nebula.
The temperature-sensitive line [OIII]~$\lambda4363$  is detected
over  11\asec (530~pc).
This region was divided into four 2\farcs9 apertures (a1 to a4), whose
positions relative to the NW HII region are shown in \fig~\ref{fig:oh}.
Table~4 lists the  measured line fluxes.
The logarithmic extinction
at \Hb was derived from the ratio of the \Ha and \Hb fluxes assuming
a stellar Balmer  absorption equivalent width of 
2\angs (e.g. Shields \& Searle 1978). 
The line ratio
$R_t \equiv F([OIII] \lambda5007 + \lambda4959) / F([OIII] \lambda4363)$
was corrected for reddening using the extinction
curve of Miller \& Matthews (1972).  
An external check of the line fluxes for the NW HII region yields 
excellent agreement with Skillman \& Kennicutt (1993), and fluxes from
slits ``a'' and ``b'' are consistent at their intersection.


Along slit~``a'', the line ratio $R_t$  shows evidence for a  significant
temperature gradient from the SW to the NE side of the nebula, which
is illustrated in the bottom panel of Figure~\ref{fig:oh}.
The photometric errors  are dominated by the 2-3\%
uncertainty of our flux calibration for bright lines and  by read noise
and photon noise in the fainter lines.
These uncertainties were combined in quadrature with the reddening uncertainty
to derive the error bars denoted by ``X'' in \fig~\ref{fig:oh}.
The error bars denoted by ``--'' include an additional term for the maximum errors introduced by
deviations from the parallactic angle.

The O/H abundance ratios shown in Figures~\ref{fig:oh}bc were computed with the IRAF
interface (Shaw \& Dufour 1995)
to the five-level atom program of De Robertis, Dufour, \& Hunt (1987).
The $T_e(O^{+})$ temperature was calculated from $T_e(O^{++})$ and the parameterization
of Pagel \et (1992) which is based on model calculations by Stasinska (1990).
The O$^{+3}$/H abundance was estimated from the HeII 4686 \angs flux, and the
contribution to the total O/H was found to be much less than the magnitude
of our error bars.


Our abundance analysis is quite sensitive to $T_e(O^{++})$.
Atmospheric dispersion shifting the  auroral
line of $O^{++}$ 
a few tenths of an arcsecond along the slit relative to the nebular lines
could cause a significant error 
in the temperature measurements.  
To test for this effect, the frames with 
$d_{\perp} < 0\farcs1$ and the frames with $d_{\parallel} < 0\farcs1$ were 
separately combined and analyzed.  The results from each set of spectra were 
similar.  However, we decided to only use the frames taken at the parallactic angle,  
$d_{\perp} < 0\farcs1$, for the abundance analysis; and
this selection criterion eliminated most of the frames along slit~b.
Unfortunately, the small aperture and steep surface brightness 
profile still make atmospheric dispersion along the slit ($d_{\parallel}$) a serious 
concern.
For these observations, the  atmospheric dispersion between 
4000 \angs and 5000 \angs is $\sles~0\farcs3$ along the slit (Filippenko 1982).
The resulting offset along the slit between the profiles of the [OIII]~4363\angs
and [OIII]~5007\angs lines  could introduce a systematic error in  $R_t$ 
as large as  $\sim 20\%$ near the inflection point in
the surface brightness profile.
This effect could introduce an artificial gradient in the measured line ratio,
but the slope would be in the opposite direction of the trend we measure.
Hence, we believe the gradient in $R_t$ is real but adopt the larger error
bars to caution the reader about a potential systematic error.
We conclude that the O/H abundance ratio in the central 11\asec (530 pc)
is witin 20\% that of the NW HII region.

\section{Summary}
\label{sec:sum}

This paper reports new kinematic evidence for
expanding, galactic-scale shells of ionized gas in  \IZw18  and  determines their
implications for the galaxy's evolution. 
We introduced a dynamical model to quantify the age and power
requirements of these superbubbles and an evolutionary synthesis model to 
associate a stellar population with them.
The mass of metals synthesized by this burst and the metallicity of the HII
regions was used to discuss the mass of interstellar gas polluted by
the starburst. The detection of oxygen emission lines from the diffuse
 gas  demonstrates the large  spatial extent  of the metal-enriched,
ionized gas.
The following conclusions can be drawn.


\begin{enumerate}
\item A supergiant shell of ionized gas extends southwestward from the galaxy.
The observed line splitting sets a firm lower limit of 30\kms on the
shell's expansion speed.  The shell could be as old as 30~Myr.

\item
On the opposite side of the NW HII region, an \Ha shell of comparable extent
is seen, and  the echellogram shows some evidence for a second
supergiant shell in this region.
These bubbles may constitute the lobes of a
supergiant bubble driven by the starburst and constricted by the 
morphology of the ambient medium to form a bipolar bubble.
Geometrical considerations suggest the shell expands at a  speed of $35 - 60$\kms.

\item 
The superbubble and starburst may be coeval.
A star formation rate of
 0.02  \msun (of 1\msun to 100\msun stars) yr$^{-1}$  
over the last $15 - 27$~Myr 
can accelerate  the shell and produce
the blue luminosity, the UBV colors, and most of the  ionizing luminosity
of IZw18.
Instantaneous burst models fail to produce enough hydrogen ionizing photons.

\item
The superbubble will probably blowout of the galaxy, and
the hot component of the ISM will escape from the galaxy in a galactic
wind.
Our analysis of the gas dynamics  supports the general picture that
galactic winds are an important process in the chemical evolution
of dwarf galaxies 
(Matteuci \& Chiosi 1983; Marconi \et 1994; Matteuci \& Tosi 1985).

\item
The current starburst will probably eject only a small fraction
of the galaxy's total gaseous mass.
Catastrophic mass loss -- proposed to transform gas-rich dwarfs into dwarf elliptical galaxies (Ferguson \& Binggeli 1994 and references therein)
 -- is unlikely from the current starburst, which may be the first in
\IZw18.
Whether this is the common outcome of wind driven mass loss from dwarf galaxies is not 
yet clear
(Meurer \et 1992; Marlowe \et 1995; Heckman \et 1995).

\item 
The detection of the superbubble establishes a timescale ($\sim$~ 15 - 27~Myr)
and spatial scale ($\sim$~900~pc) for dispersing the recently 
synthesized elements.
Oxygen emission lines are detected in the diffuse gas over this spatial scale.
A pilot study of the gas phase metal abundance suggests the
 SW side of the nebula has a lower $T(O^{++})$ and 
slightly higher oxygen abundance than the NE side.

\item
The \x emission is slightly harder and brighter than that predicted 
from the superbubble.
Higher resolution observations are needed to determine the fraction of \x
 luminosity produced by additional sources.

\end{enumerate}


\acknowledgements
I would like to thank Dave Arnett, Dave De Young, Don Garnett, 
John Salzer, and Joe Shields for valuable discussions
about this work and Sally Oey for general discussions about superbubbles.
Rob Kennicutt deserves special thanks for 
critiquing several drafts of this paper.  The constructive comments of the
referee,  Tim Heckman, improved the final presentation.
CLM acknowledges support from a NSF Graduate Fellowship.
This research was also supported in part by
NASA grant NAG5-2480 and NSF grants AST-9019150 and AST-9421145.
The NASA/IPAC Extragalactic Data Base (NED) was a useful resource for
this work.


%
%

%
%

\begin{table}
\begin{center}

TABLE~1

Properties of \IZw18
\vskip\the\baselineskip
\begin{tabular}{lll}
\tableline
\tableline
Quantity  &  Value   & Source \\
\tableline
d	& 10 Mpc     & 1 \\
B$^0_T$ & 15.8       & 2 \\
$M_B$ & -14.2 & \\
$F_{H\alpha}$ & $6.0 \pm 0.9 \times 10^{-13}$\flux & \\
$Q$ & $1.4 \pm 0.2 \times 10^{52}$~s$^{-1}$ & \\
$M_{HII}$ &   $3.9 \pm 0.4 \times 10^6 \sqrt{\frac{\epsilon}{1.5e-3}} $ & \\
$M_{HI}$ & $6.63 \pm 1.06 \times 10^7$\msun & 1 \\
(U-B)$_0^+$ & -0.88 & 3\\
(B-V)$_0^+$ & -0.01 & 3\\
\tableline
\end{tabular}
\end{center}

1) Lequeux \& Viallefond (1980)

2) Melisse \& Israel (1994)

3) Colors corrected for emission lines but not  nebular
continuum; Sudarsky (1995).

Note:  $\epsilon$ is the volume filling factor.

\end{table}

%
%
\begin{table}
\begin{center}
TABLE~2

Starburst Models\tablenotemark{a}
\end{center}
\vskip\the\baselineskip
\begin{tabular}{lccc}
\tableline
\tableline
Star Formation	History & Instantaneous &  Continuous & Continuous \\
		& Burst		&  Rate	      & Rate \\
\tableline
Age, $\tau$ (Myr) & 15	& 15	& 27 \\	
Absolute Blue Magnitude, $M_B$ & -20.4 & -14.2 & -14.2 \\
Luminosity of H Ionizing Photons, Q ($s^{-1}$) & $1.2 \times 10^{52}$  & $6.6 \times 
10^{51}$ & $5.4 \times 10^{51}$ \\
Mass of 1\msun -- 100\msun Stars, $M_*$ (\msun) & $1.4 \times 10^8$ & $3.2 \times 10^5$ &
$4.6 \times 10^5$ \\
U-B (mag) & - 0.53 & -0.78 & -0.70 \\
B-V (mag) & 0.04 & -0.03 & -0.02 \\
Kinetic Energy, $E_{in}$ (ergs) & $8.8 \times 10^{56}$ & $6.6 \times 10^{53}$ & $2.7 
\times 10^{54}$ \\
\tableline
\end{tabular}
\tablenotetext{a}{Population synthesis calculations from Leitherer \& Heckman (1995).}
\end{table}

%
%
\begin{table}
\begin{center}
TABLE~3

Gravitational Potential Model

\vskip\the\baselineskip
\begin{tabular}{cccc}
\tableline
\tableline
$r_{max}$ & r	&	$v_{\rm esc}$  & $T_{\rm esc}$ \\
\tableline
(kpc) & (pc) & (\kms) & (K) \\
\tableline
1 &200 	&	91 &1.2e5 	\\
1 & 400	&	78 &8.9e4      \\
1 & 800	&	63 &5.8e4 	\\
1 & 1000 & 57  & 4.8e4 \\
10 & 200 &125    &2.3e5\\
10 & 400  &116    &2.0e5\\
10  & 800 &106    &1.7e5\\
10 & 1000 & 103  & 1.6e5 \\
\tableline
\end{tabular}
\end{center}
Column identifications:

(1) -- Truncation radius of isothermal sphere.  The circular velocity
is fixed at 40\kms.

(2) -- Radius.

(3) -- Escape velocity at r.

(4) -- Escape temperature at r.

\end{table}
%

%
%
\begin{table}
\begin{center}
TABLE~4

MMT Spectra and Derived Emission Line Ratios\tablenotemark{a}

\vskip\the\baselineskip
\begin{tabular}{lllll}
\tableline
\tableline
Property & Region 	& 	& 	& 		\\
\tableline
 & a1& a2& a3& a4 \\
\tableline
$\lambda$3727 [OII]
			& 0.471 $\pm$ 0.017	& 0.268 $\pm$ 0.006	& 0.418 $\pm$ 0.010	& 0.497 $\pm$ 0.014			\\
H$\beta$		& 1.000 $\pm$ 0.020	& 1.000 $\pm$ 0.020	& 1.000 $\pm$ 0.020	& 1.000 $\pm$ 0.020			\\
$\lambda$4363 [OIII]	& 0.061 $\pm$ 0.005	& 0.065	$\pm$ 0.002	& 0.053 $\pm$ 0.002	& 0.037	$\pm$ 0.003		\\
$\lambda$4959 [OIII]	& 0.615 $\pm$ 0.013	& 0.716 $\pm$ 0.014	& 0.678	$\pm$ 0.014 	& 0.504 $\pm$ 0.010			\\
$\lambda$5007 [OIII]	& 1.805 $\pm$ 0.036	& 2.102 $\pm$ 0.042	& 1.998 $\pm$ 0.040	& 1.554 $\pm$ 0.031			\\
H$\alpha$		& 3.457 $\pm$0.070	& 2.883 $\pm$ 0.058	& 2.906 $\pm$ 0.058	& 2.698	$\pm$ 0.054 			\\
c(H$\beta$)\tablenotemark{b}
			& 0.29 $\pm$ 0.10	& 0.05 $\pm$ 0.10	& 0.05 $\pm$ 0.10	& -0.04 $\pm$ 0.10			\\
$R_t$\tablenotemark{c}	& 35.5 $\pm$3.1		& 42.6 $\pm$ 2.3	& 49.1 $\pm$ 3.1 	& 56.4 $\pm$5.6			\\
%
%
\tableline
\end{tabular}

\tablenotetext{a}{Emission line fluxes relative to H$\beta$; no reddening corrections have been applied. Apertures a1 - a4 are described in the text.}
\tablenotetext{b}{Logarithmic extinction at H$\beta$.}
\tablenotetext{c}{Dereddened flux ratio $R_t \equiv 
F([OIII \lambda5007 + \lambda4959) / F([OIII] \lambda4363)$}

\end{center}
\end{table}
%

%
%

%
\end{document}